\documentclass[authoryear,12pt,letter,3p]{jowarticle}
\usepackage{graphicx}
\usepackage{amsmath}
\usepackage{amssymb}
\usepackage{setspace}
\usepackage[all]{xy}
\usepackage{amsthm}
\usepackage{appendix}
\usepackage{csquotes}
\usepackage{longtable}
\usepackage{tabu}
\makeatletter
\renewcommand\section{\@startsection{section}{1}{\z@}{-3.25ex plus -1ex minus -.2ex}{1.5ex plus .2ex}{\normalsize\bf}}
\renewcommand\subsection{\@startsection{subsection}{2}{\z@}{-3.25ex plus -1ex minus -.2ex}{1.5ex plus .2ex}{\normalsize\bf}}
\renewcommand\subsubsection{\@startsection{subsubsection}{3}{\z@}{-3.25ex plus -1ex minus -.2ex}{1.5ex plus .2ex}{\normalsize\bf}}
\makeatother

\newtheorem{thm}{Theorem}

\newtheorem{prop}[thm]{Proposition}

\usepackage{multirow}

\begin{document}

\begin{frontmatter}
\title{Categories and the Foundations of Classical Field Theories}
\author{James Owen Weatherall}\ead{weatherj@uci.edu}
\address{Department of Logic and Philosophy of Science\\ University of California, Irvine, CA 92697}
\begin{abstract}
I review some recent work on applications of category theory to questions concerning theoretical structure and theoretical equivalence of classical field theories, including Newtonian gravitation, general relativity, and Yang-Mills theories.
\end{abstract}
\end{frontmatter}

\doublespacing

\section{Introduction}

There are certain questions that arise in philosophy of science---including the philosophies of particular sciences---for which it would be useful to have a formal apparatus by which to represent a scientific theory abstractly.\footnote{Of course, there is a long history of proposals aiming to do just this, beginning with \citet{CarnapAufbau}, \citet{Russell}, and \citet{Ramsey}. For overviews of this history and the current state of the field, see \citet{Lutz} and \citet{HalvorsonOxford}, and references therein.}  To this end, in the context of a sustained critique of the so-called ``semantic view'' of theories, \citet{Halvorson} has suggested that philosophers look to category theory for tools and inspiration.  The proposal is that for some purposes, it is useful to think of a scientific theory as a collection of (mathematical) models---though not, as in the semantic view, a \emph{bare} set of models, but rather as a \emph{structured} set of models, or more specifically, a \emph{category} of models.\footnote{In some cases, one might also consider topological structure \citep{Fletcher}, measure theoretic structure \citep{CurielMeasure}, etc.  The idea of using category theory to compare physical theories traces back to \citet{MacLaneGeometry}, who suggested that the Legendre transform between Hamilton and Lagrangian mechanics may be understood as a natural transformation---a result connected with recent work on the structure of classical mechanics \citep{North,Curiel,BarrettStructure}.  See also \citet{LawvereContinuum}.}

Although the present chapter develops this idea, I do not defend it.\footnote{A more direct defense can be found in \citet[this volume]{HalvorsonCategories}.  Still, I take what follows to have some probative value: in my view, fruitful application of the sort described here is the most compelling reason to develop a formal program in philosophy of science.}  Instead, my goal is to show how the idea can be put to work, by reviewing how it has been fruitfully applied to a cluster of related issues concerning symmetry, structure, and equivalence in the context of classical (i.e., non-quantum) field theories.  Contemporary interest in these issues can be traced back to \citet{Stein} and \citet{EarmanWEST}, who showed how 17th century debates concerning substantivalism and relationism in Newtonian gravitation can (and should) be understood to be about whether a classical spacetime is endowed with certain geometrical structure.\footnote{This particular thread continues: recently, \citet{Saunders}, \citet{Knox}, and \citet{WeatherallMaxwell} have considered new arguments concerning just what structure is presupposed by Newton's \emph{Principia}.}  Meanwhile, \citet{Earman+Norton} and \citet{Stachel1989} have argued that the infamous ``hole argument'' leads to puzzles about the structure of spacetime in general relativity.\footnote{For up-to-date overviews of the state of the art on the hole argument, see \citet{Pooley} and \citet{NortonSEP}.}  Recent debates concerning interpretations of Yang-Mills theory, and so-called ``gauge theories'' more generally, may be understood along similar lines \citep{Belot1,Belot2,HealeyOldest,HealeyOld,Healey,Arntzenius}.\footnote{These issues also connect up with more general concerns in philosophy of science and metaphysics, concerning (1) the role of symmetry in guiding our attempts to extract metaphysical morals from our scientific (read: physical!) theories \citep{Ismael+vanFraassen,Baker,Dasgupta} and (2) the relationship between the structural features of our theories and various forms of realism \citep{Worrall,Ladyman+Ross,French}.}

I will begin by introducing the category theory that will appear in the sequel.  Then I will show how this framework re-captures (old) intuitions about relative amounts of structure in classical spacetime theories.  This discussion will lead to classical electromagnetism and a discussion of ``gauge'' structure.  Finally, I will consider more difficult issues arising in general relativity and Yang-Mills theory.  In these last two cases, I will argue, the tools developed in the earlier parts of the paper provide new traction on thorny interpretational issues of current interest.\footnote{The material presented here draws on a number of recent papers by the author and collaborators, including \citet{WeatherallTE}, \citet{WeatherallGauge}, \citet{WeatherallHoleArg}, \citet{Rosenstock+etal}, and \citet{Rosenstock+Weatherall1,Rosenstock+Weatherall2}.}

Before proceeding, let me flag a worry.  Although some of the results I describe in the body of the paper are non-trivial, the category theory I use is elementary and, arguably, appears only superficially.  Meanwhile, there are areas of mathematics and mathematical physics closely related to the theories I discuss---synthetic differential geometry; higher gauge theory---where category theory plays a much deeper role.\footnote{For more on synthetic differential geometry, see \citet{Kock}; and for applications to physics, see \citet{MacLaneGeometry}, \citet{LawvereContinuum} and, more recently, \citet{Reyes}.  Higher gauge theory is described by \citet{Baez+Schreiber}.}   But as I said above, my goal is to review some ways in which thinking of a physical theory as a category of models bears fruit for issues of antecedent philosophical interest.  The role the category theory ends up playing is to regiment the discussion, providing the mathematical apparatus needed to make questions of theoretical structure and equivalence precise enough to settle.  The fact that fairly simple ideas bear low-hanging fruit only provides further motivation for climbing into the higher branches.

\section{Structure and Equivalence in Category Theory}\label{sec:categories}

There are many cases in which we want to say that one kind of mathematical object has more structure than another kind of mathematical object.\footnote{The ideas in this section are developed in more detail by \citet{BarrettCounting}.}  For instance, a topological space has more structure than a set.  A Lie group has more structure than a smooth manifold.  A ring has more structure than a group.  And so on.  In each of these cases, there is a sense in which the first sort of object---say, a topological space---results by taking an instance of the second sort---say, a set---and adding something more---in this case, a topology.  In other cases, we want to say that two different kinds of mathematical object have the same amount of structure. For instance, given a Boolean algebra, I can construct a special kind of topological space, known as a Stone space, from which I can uniquely reconstruct the original Boolean algebra; and vice-versa.

These sorts of relationships between mathematical objects are naturally captured in the language of category theory,\footnote{I will take for granted the basic definitions of category theory---that is, definitions of \emph{category} and \emph{functor}---but no more.  For these notions, see \citet{MacLane}, \citet{Leinster}, and other papers in this volume.}  via the notion of a \emph{forgetful functor}.  For instance, there is a functor $F:\mathbf{Top}\rightarrow\mathbf{Set}$ from the category $\mathbf{Top}$, whose objects are topological spaces and whose arrows are continuous maps, to the category $\mathbf{Set}$, whose objects are sets and whose arrows are functions.  This functor takes every topological space to its underlying set, and it takes every continuous function to its underlying function.   We say this functor is forgetful because, intuitively speaking, it forgets something: namely the choice of topology on a given set.

The idea of a forgetful functor is made precise by a classification of functors due to \citet{Baez+etal}.  This requires some machinery.  A functor $F:\mathbf{C}\rightarrow\mathbf{D}$ is said to be \emph{full} if for every pair of objects $A,B$ of $\mathbf{C}$, the map $F:\hom(A,B)\rightarrow\hom(F(A),F(B))$ induced by $F$ is surjective, where $\hom(A,B)$ is the collection of arrows from $A$ to $B$.  Likewise, $F$ is \emph{faithful} if this induced map is injective for every such pair of objects.  Finally, a functor is \emph{essentially surjective} if for every object $X$ of $\mathbf{D}$, there exists some object $A$ of $\mathbf{C}$ such that $F(A)$ is isomorphic to $X$.

If a functor is full, faithful, and essentially surjective, we will say that it forgets \emph{nothing}.  A functor $F:\mathbf{C}\rightarrow\mathbf{D}$ is full, faithful, and essentially surjective if and only if it is essentially invertible, i.e., there exists a functor $G:\mathbf{D}\rightarrow\mathbf{C}$ such that $G\circ F:\mathbf{C}\rightarrow\mathbf{C}$ is naturally isomorphic to $1_{\mathbf{C}}$, the identity functor on $\mathbf{C}$, and $F\circ G:\mathbf{D}\rightarrow\mathbf{D}$ is naturally isomorphic to $1_{\mathbf{D}}$.  (Note, then, that $G$ is also essentially invertible, and thus $G$ also forgets nothing.) This means that for each object $A$ of $\mathbf{C}$, there is an isomorphism $\eta_A:G\circ F(A)\rightarrow A$  such that for any arrow $f:A\rightarrow B$ in $\mathbf{C}$, $\eta_B\circ G\circ F(f) = f\circ \eta_A$, and similarly for every object of $\mathbf{D}$.  When two categories are related by a functor that forgets nothing, we say the categories are \emph{equivalent} and that the pair $F,G$ realizes an \emph{equivalence of categories}.

Conversely, any functor that fails to be full, faithful, and essentially surjective forgets \emph{something}.  But functors can forget in different ways.  A functor $F:\mathbf{C}\rightarrow\mathbf{D}$ forgets \emph{structure} if it is not full; \emph{properties} if it is not essentially surjective; and \emph{stuff} if it is not faithful.  Of course, ``structure'', ``property'', and ``stuff'' are technical terms in this context.  But they are intended to capture our intuitive ideas about what it means for one kind of object to have more structure (resp., properties, stuff) than another.  We can see this by considering some examples.

For instance, the functor $F:\mathbf{Top}\rightarrow\mathbf{Set}$ described above is faithful and essentially surjective, but not full, because not every function is continuous.  So this functor forgets only structure---which is just the verdict we expected.  Likewise, there is a functor $G:\mathbf{AbGrp}\rightarrow\mathbf{Grp}$ from the category $\mathbf{AbGrp}$ whose objects are Abelian groups and whose arrows are group homomorphisms to the categry $\mathbf{Grp}$ whose objects are (arbitrary) groups and whose arrows are group homomorphisms.  This functor acts as the identity on the objects and arrows of $\mathbf{AbGrp}$.  It is full and faithful, but not essentially surjective because not every group is Abelian.  So this functor forgets only properties: namely, the property of being Abelian.  Finally, consider the unique functor $H:\mathbf{Set}\rightarrow\mathbf{1}$, where $\mathbf{1}$ is the category with one object and one arrow. This functor is full and essentially surjective, but it is not faithful, so it forgets only stuff---namely all of the elements of the sets, since we may think of $\mathbf{1}$ as the category whose only object is the empty set, which has exactly one automorphism.

In what follows, we will say that one sort of object has more structure (resp. properties, stuff) than another if there is a functor from the first category to the second that forgets structure (resp. properties, stuff).  It is important to note, however, that comparisons of this sort must be relativized to a choice of functor.  In many cases, there is an obvious functor to choose---i.e., a functor that naturally captures the standard of comparison in question.  But there may be other ways of comparing mathematical objects that yield different verdicts.  For instance, there is a natural sense in which groups have more structure than sets, since any group may be thought of as a set of elements with some additional structure.  This relationship is captured by a forgetful functor $F:\mathbf{Grp}\rightarrow\mathbf{Set}$ that takes groups to their underlying sets and group homomorphisms to their underlying functions.  But any set also uniquely determines a group, known as the \emph{free group} generated by that set; likewise, functions generate group homomorphisms between free groups. This relationship is captured by a different functor, $G:\mathbf{Set}\rightarrow \mathbf{Grp}$, that takes every set to the free group generated by it and every function to the corresponding group homomorphism.  This functor forgets both structure (the generating set) and properties (the property of being a free group).  So there is a sense in which sets may be construed to have more structure than groups.

\section{Classical Spacetime Structure}\label{sec:spacetime}

To get a feel for how to apply these ideas to issues in philosophy of physics, I will begin by translating some well-understood examples into the present terms.  In particular, John \citet[Ch. 2]{EarmanWEST}, building on \citet{Stein} and others, characterizes several classical spacetime structures that have been of interest historically.  To focus on just two of the most important (the present discussion is easily extended), \emph{Galilean spacetime} consists in a quadruple $(M,t_a,h^{ab},\nabla)$, where $M$ is the manifold $\mathbb{R}^4$;\footnote{More generally, the manifolds I consider throughout the chapter are smooth, Hausdorff, paracompact, and connected.  All fields defined on these manifolds are likewise assumed to be smooth, unless I indicate otherwise.  I work in the abstract index notation, as developed by \citet{Penrose+Rindler}.  For further details on classical spacetime structure, see \citet[Ch. 4]{MalamentGR}.} $t_a$ is a one form on $M$; $h^{ab}$ is a smooth, symmetric tensor field of signature $(0,1,1,1)$, and $\nabla$ is a flat covariant derivative operator.  We require that $t_a$ and $h^{ab}$ be compatible in the sense that $t_ah^{ab}=\mathbf{0}$ at every point, and that $\nabla$ be compatible with both tensor fields, in the sense that $\nabla_a t_b=\mathbf{0}$ and $\nabla_a h^{bc}=\mathbf{0}$.

The points of $M$ represent events in space and time. The field $t_a$ is a ``temporal metric'', assigning a ``temporal length'' $|t_a\xi^a|$ to vectors $\xi^a$ at a point $p\in M$. Since $\mathbb{R}^4$ is simply connected, $\nabla_a t_b=\mathbf{0}$ implies that there exists a smooth function $t:M\rightarrow \mathbb{R}$ such that $t_a=\nabla_a t$.  We may thus define a foliation of $M$ into constant-$t$ hypersurfaces representing collections of simultaneous events---i.e., space at a time.  We assume that each of these surfaces is diffeomorphic to $\mathbb{R}^3$ and that $h^{ab}$ restricted these surfaces is (the inverse of) a flat, Euclidean, and complete metric.  In this sense, $h^{ab}$ may be thought of as a spatial metric, assigning lengths to spacelike vectors, all of which are tangent to some spatial hypersurface.  We represent particles propagating through space over time by smooth curves whose tangent vector $\xi^a$, called the \emph{4-velocity} of the particle, satisfies $\xi^a t_a =1$ along the curve.  The derivative operator $\nabla$ then provides a standard of acceleration for particles, which is given by $\xi^n\nabla_n\xi^a$.  Thus, in Galilean spacetime we have notions of objective duration between events; objective spatial distance between simultaneous events; and objective acceleration of particles moving through space over time.

However, Galilean spacetime does \emph{not} support an objective notion of the (spatial) velocity of a particle.  To get this, we move to \emph{Newtonian spacetime}, which is a quintuple $(M,t_a,h^{ab},\nabla,\eta^a)$.  The first four elements are precisely as in Galilean spacetime, with the same assumptions.  The final element, $\eta^a$, is a smooth vector field satisfying $\eta^at_a=1$ and $\nabla_a\eta^b=\mathbf{0}$.  This field represents a state of absolute rest at every point---i.e., it represents ``absolute space''.  This field allows one to define absolute velocity: given a particle passing through a point $p$ with 4-velocity $\xi^a$, the (absolute, spatial) velocity of the particle at $p$ is $\xi^a-\eta^a$.

There is a natural sense in which Newtonian spacetime has strictly more structure than Galilean spacetime: after all, it consists of Galilean spacetime plus an additional element.  As Earman observes, this judgment may be made precise by observing that the automorphisms of Newtonian spacetime---that is, its spacetime symmetries---form a proper subgroup of the automorphisms of Galilean spacetime.  The intuition here is that if a structure has \emph{more} symmetries, then there must be less structure that is preserved by the maps.\footnote{This way of comparing structure is also explored by \citet{BarrettSpacetime}, under the name SYM*.  He goes on to argue that Minkowski spacetime (see fn. \ref{Minkowski}) \emph{also} has less structure than Newtonian spacetime, but that Galilean and Minkowski spacetimes have incomparable amounts of structure.}   In the case of Newtonian spacetime, these automorphisms are diffeomorphisms $\vartheta:M\rightarrow M$ that preserve $t_a$, $h^{ab}$, $\nabla$, and $\eta^a$.  These will consist in rigid spatial rotations, spatial translations, and temporal translations (and combinations of these).  Automorphisms of Galilean spacetime, meanwhile, will be diffeomorphisms that preserve only the metrics and derivative operator.  These include all of the automorphisms of Newtonian spacetime just described, plus Galilean boosts.

It is this notion of ``more structure'' that is captured by the forgetful functor approach described above.  To recapitulate Earman, we define two categories, $\mathbf{Gal}$ and $\mathbf{New}$, which have Galilean and Newtonian spacetime as their (essentially unique) objects, respectively, and have automorphisms of these spacetimes as their arrows.\footnote{For simplicity, following \citet{Baez+etal}, I am restricting attention to groupoids, i.e., categories with only isomorphisms.  There are several ways in which one could add arrows to the categories I discuss here and in what follows, but nothing turns on how or whether one includes them.}  Then there is a functor $F:\mathbf{New}\rightarrow\mathbf{Gal}$ that takes arrows of $\mathbf{New}$ to arrows of $\mathbf{Gal}$ generated by the same automorphism of $M$.  This functor is clearly essentially surjective and faithful, but it is not full for reasons already discussed, and so it forgets only structure.  Thus the criterion of structural comparison given above perfectly recapitulates Earman's condition---and indeed, may be seen as a generalization of the latter to cases where one is comparing collections of models of a theory, rather than individual spacetimes.

To see this last point more clearly, let us move to another well-trodden example.  There are two approaches to classical gravitational theory: (ordinary) Newtonian gravitation (NG) and geometrized Newtonian gravitation (GNG), sometimes known as Newton-Cartan theory.   Models of NG consist of Galilean spacetime as described above, plus a scalar field $\varphi$, representing a \emph{gravitational potential}.  This field is required to satisfy Poisson's equation, $\nabla^a\nabla_a\varphi = 4\pi\rho$, where $\rho$ is a smooth scalar field representing the mass density on spacetime.  In the presence of a gravitational potential, massive test point particles will accelerate according to $\xi^n\nabla_n\xi^a = -\nabla^a\varphi$, where $\xi^a$ is the 4-velocity of the particle.  We write models as $(M,t_a,h^{ab},\nabla,\varphi)$.\footnote{Here and throughout, we suppress the source terms in differential equations when writing models of a theory, since such fields can generally be defined from the other fields---so, for instance, given a model $(M,t_a,h^{ab},\nabla,\varphi)$ of NG, we define an associated mass density by Poisson's equation.}

The models of GNG, meanwhile, may be written as quadruples $(M,t_a,h^{ab},\tilde{\nabla})$, where we assume for simplicity that $M$, $t_a$, and $h^{ab}$ are all as described above, and where $\tilde{\nabla}$ is a covariant derivative operator compatible with $t_a$ and $h^{ab}$.  Now, however, we allow $\tilde{\nabla}$ to be curved, with Ricci curvature satisfying the geometrized Poisson equation, $R_{ab}=4\pi\rho t_at_b$, again for some smooth scalar field $\rho$ representing the mass density.\footnote{The Ricci tensor associated with a covariant derivative operator $\nabla$ is defined by $R_{bc}=R^a{}_{bca}$, where $R^a{}_{bcd}$, is the \emph{Riemann tensor}, which is the unique tensor field such that given any point $p$ and any smooth vector field defined on a neighborhood of that point, $R^a{}_{bcd}\xi^b=-2\nabla_{[c}\nabla_{d]}\xi^a$.  The Riemann tensor vanishes iff the derivative operator is flat.}  In this theory, gravitation is not conceived as a force: even in the presence of matter, massive test point particles traverse geodesics of $\tilde{\nabla}$---where now these geodesics depend on the distribution of matter, via the geometrized Poisson equation.

There is a sense in which NG and GNG are empirically equivalent: a pair of results due to \citet{TrautmanNG} guarantee that (1) given a model of NG, there always exists a model of GNG with the same mass distribution and the same allowed trajectories for massive test point particles, and (2), with some further assumptions, vice versa \citep[see][\S 4.2]{MalamentGR}.   But in a number of influential articles, \citet{GlymourTETR, GlymourEG, GlymourTE} has argued that these are nonetheless \emph{inequivalent} theories, because of an asymmetry in the relationship just described. Given a model of NG, there is a unique corresponding model of GNG.  But given a model of GNG, there are typically many corresponding models of NG.  Thus, it appears that NG makes distinctions that GNG does not make (despite the empirical equivalence), which in turn suggests that NG has more structure than GNG.

This intuition, too, may be captured using a forget functor.  Define a category $\mathbf{NG}$ whose objects are models of NG (for various mass densities) and whose arrows are automorphisms of $M$ that preserve $t_a$, $h^{ab}$, $\nabla$, and $\varphi$; and a category $\mathbf{GNG}$ whose objects are models of GNG and whose arrows are automorphisms of $M$ that preserve $t_a$, $h^{ab}$, and $\tilde{\nabla}$.  Then there is a functor $F:\mathbf{NG}\rightarrow\mathbf{GNG}$ that takes each model of $\mathbf{NG}$ to the corresponding model given by the Trautman results, and takes each arrow to an arrow generated by the same diffeomorphism.\footnote{That this is a functor is established in \citet{WeatherallTE}.}  Then results in \citet{WeatherallTE} imply the following.
\begin{prop}\label{NGgauge}$F:\mathbf{NG}\rightarrow\mathbf{GNG}$ forgets only structure.\end{prop}

\section{Excess Structure and ``Gauge''}\label{sec:gauge}

The relationship between NG and GNG captured by Prop. \ref{NGgauge} is revealing.  As noted, there is a sense in which NG and GNG are empirically equivalent---and more, there is a sense in which the theories are capable of representing precisely the same physical situations.  And yet, NG has more structure than GNG.  This suggests that NG has more structure than is strictly necessary to represent these situations, since after all, GNG can do the same representational work with less structure.  Theories with this character are sometimes said to have ``excess structure'' and are often called \emph{gauge theories} by physicists.\footnote{For more on the term ``gauge theory'', see \citet{WeatherallGauge}.}

The archetypal example of a gauge theory is classical electromagnetism.  To be clear about the sense in which electromagnetism has excess structure, I will present two characterizations of its models.  (For simplicity, I limit attention to electromagnetism in a fixed background of Minkowski spacetime, $(M,\eta_{ab})$).\footnote{\label{Minkowski} Minkowski spacetime is a relativistic spacetime consisting of the manifold $\mathbb{R}^4$ with a flat, complete Lorentzian metric $\eta_{ab}$.  For more on relativistic spacetimes, see section \ref{sec:GR} and \citet{MalamentGR}.}  On the first formulation, which I will call EM$_1$, the fundamental dynamical quantity is a two form $F_{ab}$, known as the \emph{Faraday tensor}.  This tensor field represents the electromagnetic field at each point.\footnote{For more on how to recover electric and magnetic fields from $F_{ab}$, see \citet[\S 2.6]{MalamentGR}.}  It is required to satisfy Maxwell's equations, which may be written as $d_aF_{bc}=\mathbf{0}$ and $\nabla_aF^{ab}=J^b$, where $d$ is the exterior derivative, $\nabla$ is the Minkowski spacetime derivative operator, and $J^b$ is the \emph{charge current density} associated with any charged matter present in spacetime.  On the second formulation, EM$_2$, the fundamental dynamical quantity is a one form, $A_a$, known as the \emph{vector potential}.  The vector potential is required to satisfy $\nabla_a\nabla^a A^b = \nabla^b\nabla_a A^a + J^b$.  On the first formulation, a model of electromagnetism may be represented as an ordered triple, $(M,\eta_{ab},F_{ab})$; on the second, a model would be a triple $(M,\eta_{ab},A_a)$.

As with NG and GNG above, there is a close relationship between EM$_1$ and EM$_2$.  Given a model of EM$_2$, $(M,\eta_{ab},A_a)$, I can always construct a model of EM$_1$ that satisfies Maxwell's equations for the same charge-current density, by defining $F_{ab}=d_a A_b$.  Conversely, given a model $(M,\eta_{ab},F_{ab})$ of EM$_1$, there always exists a suitable vector potential $A_a$ such that $F_{ab}=d_aA_b$.  On both formulations, the empirical significance of the theory is taken to be captured by the associated Faraday tensor---directly for models of EM$_1$, and by the relationship $A_a\mapsto d_aA_b$ for EM$_2$.  And so, we have a clear sense in which the two formulations are empirically equivalent.

These formulations of electromagnetism are analogous to NG and GNG, as follows.  While every model of EM$_2$ gives rise to a unique model of EM$_1$, there are typically many models of EM$_2$ corresponding to any given model of EM$_1$.  Once again, this appears to have the consequence that EM$_2$ makes distinctions that EM$_1$ does not---and that these distinctions between models of EM$_2$ add nothing to the empirical success of the theory.  Again, this relationship is captured by a forgetful functor: define a category $\mathbf{EM}_1$ with models of EM$_1$ as objects and isometries of Minkowski spacetime that preserve $F_{ab}$ as arrows; and define a category $\mathbf{EM}_2$ with models of EM$_2$ as objects, and isometries of Minkowski spacetime that preserve $A_a$ as arrows.  We may then define a functor $F:\mathbf{EM}_2\rightarrow\mathbf{EM}_1$ that acts on objects via $A_a\mapsto F_{ab}=d_aA_b$, and which takes arrows to arrows generated by the same isometry.  \citet{WeatherallGauge} then proves the following.
\begin{prop}\label{EMgauge}$F:\mathbf{EM}_2\rightarrow\mathbf{EM}_1$ forgets only structure.\end{prop}

We thus have a sense in which EM$_2$ has more structure than EM$_1$---and indeed, since EM$_1$ and EM$_2$ are taken to have the same representational capacities, it would seem that EM$_2$ must have \emph{excess} structure.  This raises a question, though.  If EM$_2$ has excess structure, why do physicists use it?  The answer is purely pragmatic: vector potentials are often more convenient to work with than Faraday tensors. Moreover, the excess structure does not cause any problems for the theory, because physicists are well aware that it is there, and it is easily controlled.  In particular, if $A_a$ is a vector potential, then $A'_a=A_a+\nabla_a \psi$, for any smooth scalar field $\psi$, will be such that $d_aA_a=d_aA'_b$; moreover, $A'_a$ will satisfy the relevant differential equation for a given charge-current density just in case $A_a$ does. The map that takes $A^a$ to $A'^a$ is an example of a \emph{gauge transformation}.  Vector potentials related by gauge transformations are taken to be physically equivalent---even though they are mathematically \emph{inequivalent}, in the sense that they concern distinct fields on Minkowski spacetime.

Once we take gauge transformations into account, the two formulations are usually taken to be equivalent ways of presenting electromagnetism.  This relationship can also be captured in the language we have been developing.  In effect, that the functor $F:\mathbf{EM}_2\rightarrow\mathbf{EM}_1$ defined above fails to be full reflects the fact that, on the natural way of relating models of the theories, there are arrows ``missing'' from $\mathbf{EM}_2$.  The gauge transformations, meanwhile, may be understood as additional arrows---that is, they provide a second notion of isomorphism between models of EM$_2$ that preserves the structure that we take to have representational significance in physics.  These arrows can be added to $\mathbf{EM}_2$ to define a new category, $\mathbf{EM}'_2$, which has the same objects as $\mathbf{EM}_2$, but whose arrows $f:(M,\eta_{ab},A_a)\rightarrow(M,\eta_{ab},A'_a)$ are now pairs $f=(\chi,G_a)$, where $\chi:M\rightarrow M$ is an isometry of Minkowski spacetime, $G_a=\nabla_a\psi$ for some smooth scalar field $\psi$, and $\chi^*(A'_a)=A_a+G_a$. We may then define a new functor $F':\mathbf{EM}'_2\rightarrow\mathbf{EM}_1$, which has the same action on objects as $F$, but which acts on arrows as $(\chi,G_a)\mapsto\chi$.\footnote{That $\mathbf{EM}'_2$ is a category, and that $F'$ is a functor, is shown in \citet{WeatherallTE}.}  \citet{WeatherallTE} then proves the following.

\begin{prop}\label{EMequiv}$F':\mathbf{EM}'_2\rightarrow\mathbf{EM}_1$ forgets nothing.\end{prop}

As \citet{WeatherallTE} also shows, one can identify an analogous class of gauge transformations between models of NG.  These additional transformations, which reflect the fact that Newtonian physics cannot distinguish a state of inertial motion from one of uniform linear acceleration, have arguably been recognized as equivalences between models of Newtonian gravitation since Newton's \emph{Principia}---indeed, one may interpret Corollary VI to the Laws of Motion as describing precisely these transformations.\footnote{For more on the significance of Corollary VI, see \citet{DiSalle}, \citet{Saunders}, \citet{Knox}, and \citet{WeatherallMaxwell}.}  More, one can define an alternative category with models of NG as objects, and with these Newtonian gauge transformations included among the arrows; this new category is then equivalent to $\mathbf{GNG}$.

\section{Yang-Mills Theory and Excess Structure}\label{sec:YM}

The discussion of gauge theories in the previous section makes precise one sense in which electromagnetism and Newtonian gravitation have excess structure: there are textbook formulations of these theories that have structure that apparently plays no role in how the theories are used, in the sense that there are other formulations of the theories that have the same representational capacities, but which have less structure in the sense described by Props. \ref{NGgauge} and \ref{EMgauge}.  In many cases of interest, however, we do not have multiple formulations to work with, and we are confronted with the question of whether a particular formulation has excess structure without a comparison class.

It is here that the relationship between $\mathbf{EM}_2$ and $\mathbf{EM}'_2$ becomes especially important.  What we see in Prop. \ref{EMequiv} is that the excess structure in EM$_2$, as captured by Prop. \ref{EMgauge}, may be ``removed'' by identifying an additional class of ``gauge transformations'' that relate non-isomorphic, but physically equivalent, models of the theories.  This observation provides us with an alternative criterion for identifying when a formulation of a theory has excess structure: namely, when there are models of the theory that we believe have precisely the same representational capacities, but which are not isomorphic to one another.\footnote{This criterion is also discussed in \citet{WeatherallGauge}.}  With this criterion in mind, I will now turn to other theories that are often called gauge theories, to try to identify whether they have excess structure.  In this section, I will consider (classical) Yang-Mills theory; in the next section, general relativity.

Classically, a model of Yang-Mills theory consists in a principal connection $\omega^{\mathfrak{A}}{}_{\alpha}$ on a principal bundle $G\rightarrow P\xrightarrow{\wp} M$ with structure group $G$, over a relativistic spacetime $(M,g_{ab})$ (see section \ref{sec:GR}).\footnote{One also requires an inner product on the Lie algebra associated to $G$, but that will play no role in the following.  For more on the principal bundle formalism for Yang-Mills theory, see \citet{TrautmanYM}, \citet{Palais}, \citet{Bleecker}, \citet{Deligne+Freed}, and \citet{WeatherallYM}.  The notation used here, a variant of the abstract index notation, is explained in an appendix to \citet{WeatherallYM}.  Briefly, fraktur indices indicate valuation in a Lie algebra; lower case Greek indices label tangent vectors to the total space of a principal bundle; and lower case Latin indices label tangent vectors to spacetime.  So, for instance, $\omega^{\mathfrak{A}}{}_{\alpha}$ is a Lie algebra-valued one form on $P$, mapping tangent vectors at a point of $P$ to the Lie algebra associated to $G$.}  We will write these models as $(P,g_{ab},\omega^{\mathfrak{A}}{}_{\alpha})$.  The principal bundle $P$ may be thought of as a bundle of frames for associated vector bundles $V\rightarrow P\times_GV\xrightarrow{\pi} M$, sections of which represent distributions of matter on space-time.  The principal connection $\omega^{\mathfrak{A}}{}_{\alpha}$ determines a unique derivative operator on every vector bundle; this derivative operator is the one appearing in matter dynamics, and in this way the connection affects the evolution of matter.  Conversely, every matter field is associated with a horizontal and equivariant Lie algebra valued one form $J^{\mathfrak{A}}{}_{\alpha}$ on $P$, called the \emph{charge-current density}.  The principal connection is related to the total charge-current density on $P$ by the Yang-Mills equation, $\star\overset{\omega}{D}{}_{\alpha}\star\Omega^{\mathfrak{A}}{}_{\beta\kappa}=J^{\mathfrak{A}}{}_{\kappa}$,
where $\overset{\omega}{D}_{\alpha}$ is the exterior covariant derivative relative to $\omega^{\mathfrak{A}}{}_{\alpha}$, $\star$ is a Hodge star operator on horizontal and equivariant forms on $P$ determined by the spacetime metric $g_{ab}$ on $M$, and $\Omega^{\mathfrak{A}}{}_{\alpha\beta}$ is the curvature associated with $\omega^{\mathfrak{A}}{}_{\alpha}$.  Thus, Yang-Mills theory may be understood as a theory in which charged matter propagates in a curved space, the curvature of which is dynamically related to the distribution of charged matter.

The connection, curvature, and charge-current density may all be represented as fields on $M$ by choosing a local section $\sigma:U\subseteq M\rightarrow P$ and considering the pullbacks of these fields along the section.  The resulting fields are generally dependent on the choice of section; changes of section are known as ``gauge transformations''.  Note, however, that these gauge transformations are strongly disanalogous to the ones encountered in section \ref{sec:gauge}: they are not maps between distinct models of the theory, and so they do not indicate that there are non-isomorphic models of the theory that have the same representational capacities.  Instead, they are most naturally construed as changes of local coordinate system---or, frame field---relative to which one represents invariant fields on $P$ as fields on $M$.\footnote{One could recover a sense in which changes of section were gauge transformations in the other sense, discussed above, by stipulating that models come equipped with a preferred choice of section.  But this is not natural, for multiple reasons---the most important of which is that in general, there are no sections defined on all of $M$.}

In fact, it seems that Yang-Mills theory as just described does not have excess structure at all.  One way of emphasizing this point is to observe that classical electromagnetism is a Yang-Mills theory in the present sense, which means we can compare EM$_1$ and EM$_2$ with a third formulation of the theory, EM$_3$.  EM$_3$ is a Yang-Mills theory with structure group $U(1)$.  The Lie algebra associated with this group is $\mathbb{R}$, and so Lie algebra valued forms are just real valued, which means we can drop the index $\mathfrak{A}$ from the curvature and connection forms.  Continuing as above, we limit attention to Minkowski spacetime.  Thus a model of the theory is a triple $(P,\eta_{ab},\omega_{\alpha})$, where $\omega_{\alpha}$ is a principal connection on $U(1)\rightarrow P\xrightarrow{\wp} M$, the unique (necessarily trivial) $U(1)$ bundle over Minkowski spacetime.  Given such a model, we may generate models of EM$_1$ or EM$_2$ by choosing a (global) section $\sigma:M\rightarrow P$, and defining a Faraday tensor $F_{ab}=\sigma^*(\Omega_{\alpha\beta})$ or a vector potential $A_a=\sigma^*(\omega_{\alpha})$, respectively.  Since $U(1)$ is Abelian, $F_{ab}$ is independent of the choice of section; $A_a$, however, depends on the section, with different sections producing different (gauge related) vector potentials.  If $\omega_{\alpha}$ satisfies the Yang-Mills equation for a charge-current density $J_{\alpha}$, then $A_a$ and $F_{ab}$ will satisfy their respective equations of motion for a field $J_a=\sigma^*(J_{\alpha})$.

We now define a category $\mathbf{EM}_3$ whose objects are models of EM$_3$ and whose arrows are principal bundle isomorphisms that preserve both $\omega_{\alpha}$ and the Minkowski metric $\eta_{ab}$.  There is also a natural functor $F:\mathbf{EM}_3\rightarrow\mathbf{EM}_1$, which acts on objects as $(P,\eta_{ab},\omega_{\alpha})\mapsto (M,\eta_{ab},d_a\sigma^*(\omega_{\alpha}))$ (for any global section $\sigma$) and on arrows as $(\Psi,\psi)\mapsto \psi$.  \citet{WeatherallGauge} then proves the following result.
\begin{prop}\label{UGProp}$F:\mathbf{EM}_3\rightarrow\mathbf{EM}_1$ forgets nothing.\end{prop}
It follows that, by the criterion of comparison we have been considering, the principal bundle formalism has the same amount of structure as EM$_1$, and less structure than EM$_2$.  This is despite the fact that the dynamical variable of the theory is the \emph{connection} on $P$, which is analogous---via a choice of section---to the vector potential.  The reason the equivalence holds is that given any diffeomorphism that preserves the Faraday tensor, there is a unique corresponding principal bundle automorphism that also preserves $\omega_{\alpha}$, in effect by systematically relating the possible sections of $P$.

These variants of electromagnetism on Minkowski spacetime are ultimately toy examples.  That said, there is another issue in the neighborhood that has been a locus of recent debate.  It concerns the relationship between the formalism for Yang-Mills theory just described---the so-called ``fiber bundle formalism''---and a formalism known as the ``holonomy formalism'' or ``loop formalism''.  Each of these is often associated with an ``interpretation'' of Yang-Mills theory, though I will not discuss those interpretations.\footnote{In particular, see \citet{Healey} and \citet{Arntzenius} for discussions of the interpretive options related to these formalisms; see also \citet{Rosenstock+Weatherall2}.}  It is sufficient to note that one theme in these debates is the claim that the holonomy formalism posits (or requires) less structure than the fiber bundle formalism.

The idea behind the holonomy formalism for Yang-Mills theory is as follows.\footnote{For details, see \citet{Rosenstock+Weatherall1}; for further background, see \citet{Barrett}, \citet{Loll}, and \citet{Gambini+Pullin}.}  Given a fiber bundle model of a Yang-Mills theory, $(P,g_{ab},\omega^{\mathfrak{A}}{}_{\alpha})$, and a fixed point $u\in P$, we may define a map $H:L_{\wp(u)} \rightarrow G$, where $L_{\wp(u)}$ is the collection of piecewise smooth closed curves $\gamma:[0,1]\rightarrow M$ originating (and ending) at $\wp(u)$. This map assigns to each curve $\gamma\in L_{\wp(u)}$ the element $g\in G$ such that $\gamma_{u}(1)=\gamma_{u}(0)g$, where $\gamma_u:[0,1]\rightarrow P$ is the (unique) horizontal lift of $\gamma$ through $u$, relative to $\omega^{\mathfrak{A}}{}_{\alpha}$.\footnote{The horizontal lift of a curve $\gamma:[0,1]\rightarrow M$ through $u\in\wp^{-1}[\gamma(0)]$ is the unique curve $\gamma_{u}:[0,1]\rightarrow P$ such that $\pi\circ\gamma_u=\gamma$ and $\omega^{\mathfrak{A}}{}_{\alpha}\xi^{\alpha}=\mathbf{0}$ along the curve, where $\xi^{\alpha}$ is the tangent to $\gamma_u$. See \citet[Ch. 2]{Kobayashi+Nomizu}.}  The group element $H(\gamma)$ is known as the \emph{holonomy} of $\gamma$.  One usually interprets $\gamma_u(1)\in P$ to be the \emph{parallel transport} of $u$ along $\gamma$, relative to the principal connection $\omega^{\mathfrak{A}}_{\alpha}$; thus, the holonomies associated with a connection encode information about the parallel transport properties of $\omega^{\mathfrak{A}}{}_{\alpha}$.

It is arguably the case that the holonomies of a principal connection contain all of the empirically significant data associated with a fiber bundle model of the theory.\footnote{See \citet{Healey} for a defense of this claim.  It is usually motivated by arguing that the empirical significance of a Yang-Mills theory concerns only interference effects exhibited by quantum particles.}  The holonomy formalism attempts to characterize Yang-Mills theory directly with holonomy data.  It does so via a \emph{generalized holonomy map} $H:L_x\rightarrow G$, which simply assigns group elements to closed curves without ever mentioning a principal bundle.  (Not any map will do; the properties required of a generalized holonomy map are described in \citet{Barrett} and \citet{Rosenstock+Weatherall1}.)  Many commentators have had the intuition that this approach is more parsimonious than the fiber bundle formalism, since it posits just the structure needed to encode the possible predictions of the theory, without any geometrical superstructure.

But can this thesis of relative parsimony be made precise?  It is not clear that it can be.  In fact, the methods described here give a strikingly different answer.  For any given Yang-Mills theory, circumscribed by some fixed choice of structure group $G$, we may define two categories of models, associated with each of these formalisms.  The first, $\mathbf{PC}$, is a generalization of $\mathbf{EM}_3$: the objects are fiber bundle models $(P,g_{ab},\omega^{\mathfrak{A}}{}_{\alpha})$, with structure group $G$; and the arrows are principal bundle isomorphisms that preserve the connection and metric.  The second, $\mathbf{Hol}$, corresponds to the holonomy formalism.  Here the objects are \emph{holonomy models}, $(M,g_{ab},H:L_x\rightarrow G)$, where $H$ is a generalized holonomy map, and the arrows are \emph{holonomy isomorphisms}, which are maps that preserve the metric and holonomy structure.  (These are somewhat subtle, and are described in detail in \citet{Rosenstock+Weatherall1}.)  There is a functor $F:\mathbf{Hol}\rightarrow\mathbf{PC}$ that takes every holonomy model $(M,g_{ab},H)$ to a fiber bundle model that gives rise to the holonomies $H$, with appropriately compatible action on arrows.  \citet{Rosenstock+Weatherall1} then prove the following result.
\begin{prop}$F:\mathbf{Hol}\rightarrow\mathbf{PC}$ forgets nothing.\end{prop}
In other words---philosophers' intuitions notwithstanding---the formalisms underlying holonomy and fiber bundle interpretations have precisely the same amounts of structure, relative to a natural standard of comparison between models of the theories.\footnote{Space constraints prevent further elaboration on this point; see \citet{Rosenstock+Weatherall2} for further details.}

\section{General Relativity, Einstein Algebras, and the Hole Argument}\label{sec:GR}

Finally, I will return to spacetime physics, to discuss the methods above in the context of general relativity.  I will briefly discuss two related topics.  The first concerns the so-called ``hole argument'' of \citet{Earman+Norton} and \citet{Stachel1989}; the second concerns a proposal for an alternative to the standard formalism of general relativity originally due to \citet{GerochEA}, and later championed by \citet{Earman1977, Earman1978, Earman1979, Earman1986, Earman1989, EarmanWEST}.\footnote{For more on the hole argument, see \citet{EarmanWEST} and \citet{Pooley}; see also \citet{WeatherallHoleArg}, which extends the arguments given here.  \citet{StachelLR} also provides a detailed discussion of the hole argument in the context of category theory, though the perspective is somewhat different from that adopted here.}

To begin, the models of relativity theory are \emph{relativistic spacetimes}, which are pairs $(M,g_{ab})$ consisting of a 4-manifold $M$ and a smooth, Lorentz-signature metric $g_{ab}$.\footnote{My notational conventions again follow \citet{MalamentGR}.  In particular, I work with a signature $(1,-1,-1,-1)$ metric.}  The metric represents geometrical facts about spacetime, such as the spatiotemporal distance along a curve, the volume of regions of spacetime, and the angles between vectors at a point.  It also characterizes the motion of matter: the metric $g_{ab}$ determines a unique torsion-free derivative operator $\nabla$, which provides the standard of constancy in the equations of motion for matter.  Meanwhile, geodesics of this derivative operator whose tangent vectors $\xi^a$ satisfy $g_{ab}\xi^a\xi^b>0$ are the possible trajectories for free massive test particles, in the absence of external forces.  The distribution of matter in space and time determines the geometry of spacetime via Einstein's equation, $R_{ab}-\frac{1}{2} R g_{ab} = 8\pi T_{ab}$, where $T_{ab}$ is the \emph{energy-momentum tensor} associated with any matter present, $R_{ab}$ is the Ricci tensor, and $R=R^a{}_a$.  Thus, as in Yang-Mills theory, matter propagates through a curved space, the curvature of which depends on the distribution of matter in spacetime.

The most widely discussed topic in the philosophy of general relativity over the last thirty years has been the hole argument, which goes as follows.  Fix some spacetime $(M,g_{ab})$, and consider some open set $O\subseteq M$ with compact closure.  For convenience, assume $T_{ab}=\mathbf{0}$ everywhere. Now pick some diffeomorphism $\psi:M\rightarrow M$ such that $\psi_{|M-O}$ acts as the identity, but $\psi_{|O}$ is not the identity.  This is sufficient to guarantee that $\psi$ is a non-trivial automorphism of $M$.  In general, $\psi$ will not be an isometry, but one can always define a new spacetime $(M,\psi^*(g_{ab}))$ that is guaranteed to be isometric to $(M,g_{ab})$, with the isometry realized by $\psi$.  This yields two relativistic spacetimes, both representing possible physical configurations, that agree on the value of the metric at every point outside of $O$, but in general disagree at points within $O$.  This means that the metric outside of $O$, including at all points in the past of $O$, cannot determine the metric at a point $p\in O$.  Thus, \citet{Earman+Norton} argue, general relativity, as standardly presented, faces a pernicious form of indeterminism.  To avoid this indeterminism, one must become a \emph{relationist} and accept that ``Leibniz equivalent'', i.e., isometric, spacetimes represent the same physical situations.  The person who denies this latter view---and thus faces the indeterminism---is dubbed a \emph{manifold substantivalist}.

One way of understanding the dialectical context of the hole argument is as a dispute concerning the correct notion of equivalence between relativistic spacetimes.  The manifold substantivalist claims that isometric spacetimes are \emph{not} equivalent, whereas the relationist claims that they are.  In the present context, these views correspond to different choices of arrows for the categories of models of general relativity.  The relationist would say that general relativity should be associated with the category $\mathbf{GR}_1$, whose objects are relativistic spacetimes and whose arrows are isometries.  The manifold substantivalist, meanwhile, would claim that the right category is $\mathbf{GR}_2$, whose objects are again relativistic spacetimes, but which has only identity arrows.  Clearly there is a functor $F:\mathbf{GR}_2\rightarrow\mathbf{GR}_1$ that acts as the identity on both objects and arrows and forgets only structure.  Thus the manifold substantivalist posits more structure than the relationist.

Manifold substantivalism might seem puzzling---after all, we have said that a relativistic spacetime is a Lorentzian manifold $(M,g_{ab})$, and the theory of pseudo-Riemannian manifolds provides a perfectly good standard of equivalence for Lorentzian manifolds \emph{qua} mathematical objects: namely, isometry.  Indeed, while one may stipulate that the objects of $\mathbf{GR}_2$ are relativistic spacetimes, the arrows of the category do not reflect that choice.  One way of charitably interpreting the manifold substantivalist is to say that in order to provide an adequate representation of \emph{all} the physical facts, one actually needs \emph{more} than a Lorentzian manifold.  This extra structure might be something like a fixed collection of labels for the points of the manifold, encoding which point in physical spacetime is represented by a given point in the manifold.\footnote{A similar suggestion is made by \citet{Stachel1993}.}  Isomorphisms would then need to preserve these labels, so spacetimes would have no non-trivial automorphisms.  On this view, one might \emph{use} Lorentzian manifolds, without the extra labels, for various purposes, but when one does so, one does not represent all of the facts one might (sometimes) care about.

In the context of the hole argument, isometries are sometimes described as the ``gauge transformations'' of relativity theory; they are then taken as evidence that general relativity has excess structure.  But as I argued in section \ref{sec:YM}, one can expect to have excess structure in a formalism only if there are models of the theory that have the same representational capacities, but which are \emph{not} isomorphic as mathematical objects.  If we take models of GR to be Lorentzian manifolds, then that criterion is not met: isometries are precisely the isomorphisms of these mathematical objects, and so general relativity does \emph{not} have excess structure.

This point may be made in another way.  Motivated in part by the idea that the standard formalism has excess structure, Earman has proposed moving to the alternative formalism of so-called Einstein algebras for general relativity, arguing that Einstein algebras have less structure than relativistic spacetimes.\footnote{For details, see \citet{Rosenstock+etal}.  In many ways, the arguments here are reminiscent of those of \citet{Rynasiewicz}.}  In what follows, a \emph{smooth $n-$algebra} $A$ is an algebra isomorphic (as algebras) to the algebra $C^{\infty}(M)$ of smooth real-valued functions on some smooth $n-$manifold, $M$.\footnote{These may also be characterized in purely algebraic terms \citep{Nestruev}.}  A \emph{derivation} on $A$ is an $\mathbb{R}$-linear map $\xi:A\rightarrow A$ satisfying the Leibniz rule, $\xi(ab)=a\xi(b) + b\xi(a)$.  The space of derivations on $A$ forms an $A$-module, $\Gamma(A)$, elements of which are analogous to smooth vector fields on $M$.  Likewise, one may define a dual module, $\Gamma^*(A)$, of linear functionals on $\Gamma(A)$.  A \emph{metric}, then, is a module isomorphism $g:\Gamma(A)\rightarrow\Gamma^*(A)$ that is symmetric in the sense that for any $\xi,\eta\in\Gamma(A)$, $g(\xi)(\eta)=g(\eta)(\xi)$.  With some further work, one can capture a notion of signature of such metrics, exactly analogously to metrics on a manifold.  An \emph{Einstein algebra}, then, is a pair $(A,g)$, where $A$ is a smooth $4-$algebra and $g$ is a Lorentz signature metric.

Einstein algebras arguably provide a ``relationist'' formalism for general relativity, since one specifies a model by characterizing (algebraic) relations between possible states of matter, represented by scalar fields.  It turns out that one may then reconstruct a unique relativistic spacetime, up to isometry, from these relations by representing an Einstein algebra as the algebra of functions on a smooth manifold.  The question, though, is whether this formalism really eliminates structure.  Let $\mathbf{GR}_1$ be as above, and define $\mathbf{EA}$ to be the category whose objects are Einstein algebras and whose arrows are algebra homomorphisms that preserve the metric $g$ (in a way made precise by \citet{Rosenstock+etal}).  Define a \emph{contravariant} functor $F:\mathbf{GR}_1\rightarrow\mathbf{EA}$ that takes relativistic spacetimes $(M,g_{ab})$ to Einstein algebras $(C^{\infty}(M),g)$, where $g$ is determined by the action of $g_{ab}$ on smooth vector fields on $M$, and takes isometries $\psi:(M,g_{ab})\rightarrow (M',g'_{ab})$ to algebra isomorphisms $\hat{\psi}:C^{\infty}(M')\rightarrow C^{\infty}(M)$, defined by $\hat{\psi}(a)=a\circ\psi$.\footnote{A contravariant functor is one that takes arrows $f:A\rightarrow B$ to arrows $F(f):F(B)\rightarrow F(A)$.  The classification described above carries over to contravariant functors, though two categories related by a full, faithful, essentially surjective contravariant functor are said to be \emph{dual}, rather than \emph{equivalent}.}  \citet{Rosenstock+etal} prove the following.

\begin{prop}$F:\mathbf{GR}_1\rightarrow \mathbf{EA}$ forgets nothing.\end{prop}

\section{Conclusion}

I have reviewed several cases in which representing a scientific theory as a category of models is useful for understanding the structure associated with a theory.  In the context of classical spacetime structure, the category theoretic machinery merely recovers relationships that have long been appreciated by philosophers of physics; these cases are perhaps best understood as litmus tests for the notion of ``structure'' described here. In the other cases, the new machinery appears to do useful work.  It helps crystalize the sense in which EM$_2$ and NG have excess structure, in a way that clarifies an important distinction between these theories and other kinds of gauge theories, such as Yang-Mills theory and general relativity.  It also clarifies the relationship between various formulations of physical theories that have been of interest to philosophers because of their alleged parsimony.  These results seem to reflect real progress in our understanding of these theories---progress that apparently required the basic category theory used here.  One hopes that these methods may be extended further---perhaps to issues concerning the relationships between algebras of observables and their representations in quantum field theory and the status of dualities in string theory.\footnote{For more on the former, see \citet{Ruetsche} and references therein.}

\section*{Acknowledgments}
This material is based upon work supported by the National Science Foundation under Grant No. 1331126.  Thank you to Thomas Barrett, Hans Halvorson, Ben Feintzeig, Sam Fletcher, David Malament, Sarita Rosenstock, and two anonymous referees for many discussions related to this material, and to Thomas Barrett, Ben Feintzeig, JB Manchak, and Sarita Rosenstock for comments on a previous draft.
\singlespacing
\bibliography{gaugeMaster}
\bibliographystyle{elsarticle-harv}

\end{document}